# Silicon Photonics-based Heterodyne Interferometric Imager for free-space imaging

*Humphry Chen\*, Mingye Fu, Shun-Hung Lee, Shelbe Timothy, Lawrence Shing, Gopal Vasudevan, Tony Kowalczyk, Neal Hurlburt, and Sung-Joo Ben Yoo*

Department of Electrical and Computer Engineering, University of California Davis, One Shields Ave, Davis, California 95616.

Lockheed Martin Advanced Technology Center, Lockheed Martin, 3251 Hanover St. Palo Alto, California 94304

ABSTRACT: This paper reports on the design, fabrication, and demonstration of a silicon photonics based heterodyne interferometric imaging system. The photonic integrated circuit (PIC) can perform one-dimensional spectroscopy for unique input spectrums using a single baseline within its 91 available baselines. The PIC uses polarization diversifying gratings to separate incoming light into two distinct polarizations, an on-chip 2x4 optical hybrid, and a strong local oscillator (LO) to perform the heterodyne measurements. The optical hybrids combine the input signals with the LO and splitting them into 2 components pairs for phase sensitive measurements. Furthermore, the PIC can perform 2-D image reconstruction by combining many baseline pairs to measure the visibility of a simple target. These demonstrations show the PIC's capabilities for 1-D spectroscopy and 2-D imaging applications.

**Introduction**

High resolution stellar images are crucial for understanding the scientific phenomena of stars within the universe. A standard spectropolarimeter system is widely used to observe scientific phenomena such as Zeeman splitting. In order to observe the light, a polarization modulator is used to help isolate specific polarization components during measurement. From there, the light is fed into a spectrometer system to separate different wavelengths of light prior to detection. The Daniel K. Inouye Solar Telescope (DKIST) [T. Rimmele et al 2020] and the Hinode Solar Optical Telescope (SOT) have large optical trains that process the light prior to capture within the various measurement instruments within the system. However, these systems require several large optics and electronics to maintain and calibrate the system for spectro-polarimetry. In addition, to achieve the specific spatial resolution requirements, the external stabilization instruments become numerous and increase the SWaP of the system.

The DKIST system uses several fold mirrors to redirect light within the instrument. To stabilize the signal, active optics and electronics are used to manipulate the position and alignment of the mirrors while hundreds of air jets and liquid cooling systems are used to prevent distortions in the mirrors by cooling each mirror to 0°C. This contributes to the large size of the Gregorian Optical Station (GOS) which is a combined size of 4 m x 2 m x 1 m. In the SOT, several fold mirrors are used to redirect light and filter the signal towards the detectors. Heat dumps are used to remove excess heat generated by the incoming light, preserving the reflectivity of the mirrors. A polarization modulator is also included in the system to help capture filtergrams depending on the position of the modulator. This modulator rotates, allowing for sequential capture of information of different polarization states.

Interferometric imaging is an alternative to direct imaging by using many smaller sub-apertures in place of a large primary mirror [J.W. Goodman, 2000]. An interferometer array creates fringes from interfering signals between aperture pairs and reconstructing an image using van Cittert-Zernike theorem. Within this, spectral information can be extracted by directly examining the fringes. In addition, techniques such as heterodyne mixing can be used to improve signal strength and extract phase information. The Berkeley Infrared Spatial Interferometer (ISI) is an example of a heterodyne interferometer system [M. Bester, 1994]. The heterodyne detection scheme helps to sample input signals using a strong tunable reference laser and down-converts their signals into radio-frequency. However, the challenges of these interferometry systems are the scalability of the array to cover the Fourier space (UV-plane) and stability of the entire system.

Photonic-Integrated-Circuits (PICs) utilizing Silicon Photonics (SiPh) platforms are an attractive platform to address the SWaP challenges interferometric imaging and spectroscopic systems face [Jiang 2001, A. Arbabi (2013)]. PICs incorporate optical components to greatly reduce the size of the apertures within the interferometry system. The platform provides simple methods of scalability within a compact area, and commercial foundries offer process design kits (PDKs). These PDKs enable designers to easily utilize their design rules for nanoscale photonic designs and use pre-designed devices in the PDK libraries to greatly reduce design cycles and achieve predictable outcomes. Unlike electronic integrated circuits (EICs), PICs are not affected by thermal (Johnson) noise of resistors or electromagnetic interference (EMI) noises from neighboring signals. SiPh PICs utilize silicon-on-insulator wafers which provide thermo-mechanically stable platforms while exploiting that already well-established CMOS packaging ecosystem [G.N. Tzintzarov, 2021]. Our previous demonstration of interferometric imaging, the SPIDER interferometry system demonstrated a silicon nitride based PIC platform to perform 12 baseline and 18 spectral bin interferometric imaging [T. Su.2018]. PIC-based imaging can be extended to the studies of the Sun and stars. In particular, Doppler shifts of plasma provides information of the movement or rotation of the star, while Zeeman splitting of the atomic spectral lines offer the strength of the magnetic field.

The Zeeman effect is characterized by the splitting of a spectral line in the presence of a magnetic field, where the frequency shift of the lines are proportional to the strength of the magnetic field component along the line of sight. The Doppler shift is observed from the frequency shift of a spectral line due to the movement of an object along the line of sight, like the rotation of the sun. To distinguish between these two phenomena, the imaging system would need to measure the frequency shifts of circularly polarization of light. Each polarization of light will frequency shift in different directions for the Zeeman effect, while all polarizations of light will frequency shift in the same direction for Doppler shifts. Systems such as the Helioseismic and Magnetogram Imager (HMI) [P.H. Scherrer, 2012] and the Vector Spectromagnetograph (VSM) system in the Synoptic Optical Long-Term Investigations of the Sun (SOLIS) [C. U. Keller, 2003] capture magnetograms with bulky optical systems They can also measure the transverse fields by observing the linear polarization of the unsplit Zeeman signal. Often, only one polarization and wavelength of light is captured at a time, making measurement times increase as the system has to recalibrate and recapture the image. The first heterodyne spectroscopic measurements of sunspots have also demonstrated using a tunable diode laser around 8 microns [Glenar, 1983].

To image all of the features of the sun, we require high spatial resolution imaging capabilities while also being able to process the solar signal in real-time. Traditional system are typically too bulky, too power-hungry, or too costly to create a system that meets the exacting requirements and future instruments that scale beyond current requirements. In order to continue to achieve higher resolution systems, we want to pursue low SWaP methods to enable continuous scalability in multiple dimensions.

We can design a photonics-based interferometric imaging spectrograph and capture spectral information across multiple polarizations using heterodyne detections. By using heterodyne detection, we can amplify weak optical inputs and perform narrowband measurements of spectral features using a tunable reference laser. Frequency shifts from Zeeman splitting and doppler shifting can be observed as the reference laser is tuned, while multiple polarizations can be observed with unique optical design. In our previous work [H. Chen, 2024], we have demonstrated small-scale heterodyne interferometer design using a single baseline and a combination of fiber inputs into the PIC. The fiber inputs in that experiment allow for direct coupling of light into the gratings of the baseline. However, this method only allows for the coupling of light into two specific apertures, limiting the applications to single baseline measurements for specific inputs. We replace this input scheme with a free-space source, allowing the PIC to utilize various apertures across the PIC area.

In this paper, we demonstrate a new imaging system on PIC using the MICRO principle, a silicon photonics-based heterodyne interferometric imaging PIC with 14 apertures. The design is based on a 3-layer silicon nitride platform used to create 2-D polarization diversifying gratings used for observing two distinct polarizations of light. The PIC is used to capture unique input spectra for 1-D spectroscopy using the shortest baseline. Then, we use the baselines across the PIC surface to perform 2-D interferometric imaging of point-source functions, collimated and directed to the PIC surface. The PIC incorporates polarization diversifying gratings, on-chip thermal phase shifters, and on-chip 2x4 optical hybrids to enable phase sensitive detections of multiple polarizations of light.

**Heterodyne Imaging**

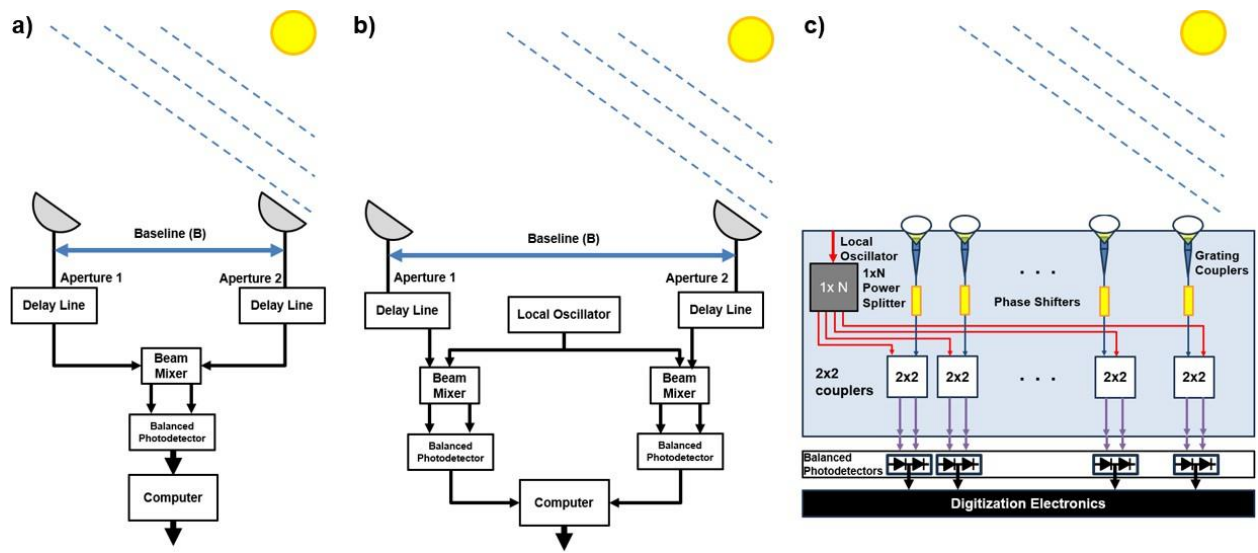

Figure 1. Standard interferometry system scheme with two apertures (a) compared to a heterodyne interferometry system with two apertures (b). A working principle design for a heterodyne interferometry PIC.

The MICRO interferometer design adds polarization diversified heterodyne imaging methods on the optical interferometry techniques of the SPIDER interferometer. The SPIDER interferometer uses two apertures, also called a baseline, to capture light from a distant source. The captured signals are combined and mixed prior to detection using a pair of detectors, shown in Figure 1. The direct combination of these signals creates a correlated signal needed to perform interferometry and capture image information as defined by the van-Cittert Ziernike theorem, given by the equation below:

$$V = \int_0^\infty f_1(t) f_2(t - \tau) d\tau$$

Where f₁ and f₂ are the signals of two apertures in a baseline and V is the visibility of the object being observed. The visibility of the object is defined as the inverse-Fourier transform of the image, allowing us to indirectly sample the image with two smaller apertures.

The MICRO interferometer builds upon the SPIDER design by introducing a reference laser, also known as a local oscillator (LO) to perform heterodyne mixing of each aperture's signals prior to combining the baseline signals together and detection using a balanced photodetector, shown in Figure 1. In the MICRO design, the captured photocurrent signal of one aperture is given by the theoretical equation below:

$$I = 2R\sqrt{P_S}\sqrt{P_{LO}}\cos(\Delta\omega_{IF}t - \phi_S - \phi_{LO} + \Delta x)$$

Here, $P_S$ is the optical power captured by the aperture, $P_{LO}$ is the optical power of the LO that's mixed with the input signal, and R is the responsivity of the detectors. The $\Delta\omega_{IF}$ term refers to the intermediate frequency that is created from the difference between the input signal's frequency and the LO's frequency. $\Delta\Phi_S$ is the inherent phase of the input signal as it travels to the aperture and $\Delta\Phi_S$ is the phase of the LO as it travels through the PIC. Lastly, $\Delta x$ refers to the phase delay term that is induced from the phase shifters on the LO. The photocurrent of one aperture is proportional to the strength of the LO which can help improve the signal-to-noise ratio of weak input signals, such as stellar light. By multiplying the photocurrents at the two apertures, the total current is given by:

$$I = 4RP_S P_{LO}\cos(\Delta\phi_1 - \Delta\phi_2 + \Delta x)$$

$\Delta\Phi_1$ and $\Delta\Phi_2$ are the total phases of the signals coming from each aperture of the observed baseline, which include the phases of the LO and signal for each aperture. $\Delta x$ is given as the phase difference induced by a combination of phase shifters between the two arms of a baseline. This phase difference can be used to nullify the other phase terms as well as observe the maximum and minimum current generated by the baseline. This is useful for determining the amplitude of the visibility captured by the baseline, which is defined as

$$abs(V) = abs(\frac{I_{max} - I_{min}}{I_{max} + I_{min}})$$

By sampling across a wide variety of baseline lengths and positions, we can use the van Cittert-Ziernike theorem to generate the Fourier Transform of a 2-D image of an input source. From there, we can take the inverse Fourier Transform to reconstruct the input source distribution. In Figure 1c, the $Si_3N_4$ PIC uses on-chip gratings to couple light and separate the signals into two polarizations. The gratings are along the curve of a semicircle and provides the longest baseline distance, $B_{max}$, along the diameter of the semicircle. Each aperture is irregularly spaced to provide as many unique baselines, with two apertures placed as close as possible to create the minimum baseline distance, $B_{min}$. The signals of each grating are coupled into the waveguides and combines the signal with the LO in a 2 x 4 optical hybrid. A set of balanced photodetectors are used to measure the output signals and are correlated off the PIC. Once correlated, the interferogram helps to provide 1-D information from the baseline, which is summed with the information from other baselines to create 2-D Fourier transformed images.

**Device Design**

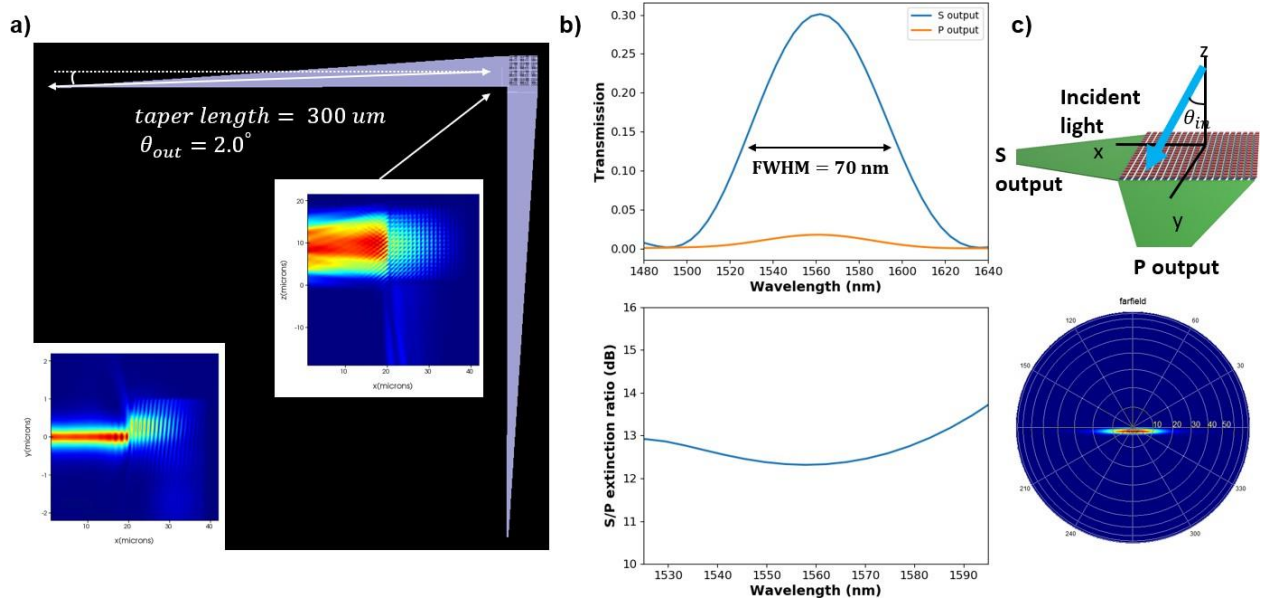

Figure 2. a) The grating design with FDTD mode field simulations discussing the polarization isolation given a single polarized light input. b) Simulated graphs for the full width half maximum (FWHM) of the gratings and the extinction ratio between the *S*- and *P*- polarizations, which is simulated to be 13 dB. c) Diagram of the grating with the incident angle and a graph of the incident angle, which is simulated to be approximately 2°.

We design, fabricate, and demonstrate a *Si₃N₄* PIC to create a scalable, compact platform. For the PIC, we use the polarization diversifying grating as a basis for setting the layer specifications. In this design, we utilize three *Si₃N₄* waveguide layers that have a thickness of 150 nm each. The waveguide layers are separated with 50 nm of *SiO₂*, acting as a cladding layer between each waveguide layer. Each waveguide layer is used in creating the polarization diversifying grating. The grating region is composed of 650 nm x 650 nm squares with a pitch of 1.082 μm, used to form a region that's a 10 μm x 10 μm. Each grating layer is offset by 250 nm horizontally in the x and y axes, allowing for each layer to be offset from each other to promote coupling of the light down into the waveguide layers and separation of the two polarizations of light. The result of the grating design is that the incident angle is 2° from the normal, angled 45° from the corner of the grating array as shown in Figure 2, allowing for near-vertical coupling of light. Once the light is separated, it is guided into a taper which guides and compresses the light to propagate in a 2 μm wide waveguide. The grating has a full-width half max (FWHM) of 70 nm centered at 1560 nm and has a transmission value of approximately 0.3, or about 30% transmission, at the output of the taper for the desired polarization. This translates to approximately 5.22 dB of loss from input to output.

The PIC is designed to have 14 apertures spaced irregularly along the arc of a semicircle with a diameter of 22 mm. A common local oscillator (LO) is externally acquired by edge coupling and promptly split using a 1 x 16 splitter. Each aperture uses the polarization diversifying grating to split incoming light into its two polarized components. Each component of the input signal is routed through the PIC and directed to the on-chip 2 x 4 optical hybrid along with the shared LO. The optical hybrid is designed with a built-in 2 x 2 coupler for one of the incoming signals, which would give a 90° phase shift due to the properties of the device [Y. Nasu, 2011]. The captured signals can be detected using balanced photodetectors to create an in-phase (I) and quadrature phase (Q) component, which are separated by 90° in phase. By using the principles of coherent detection, we can extract the phase information of the incoming signal, which can be used for examining features beyond amplitude. Within the optical hybrid, thermal phase shifters are added to the design to enable adjustments to the 90° phase shift as well as enable phase tuning of the PIC for generating fringe data.

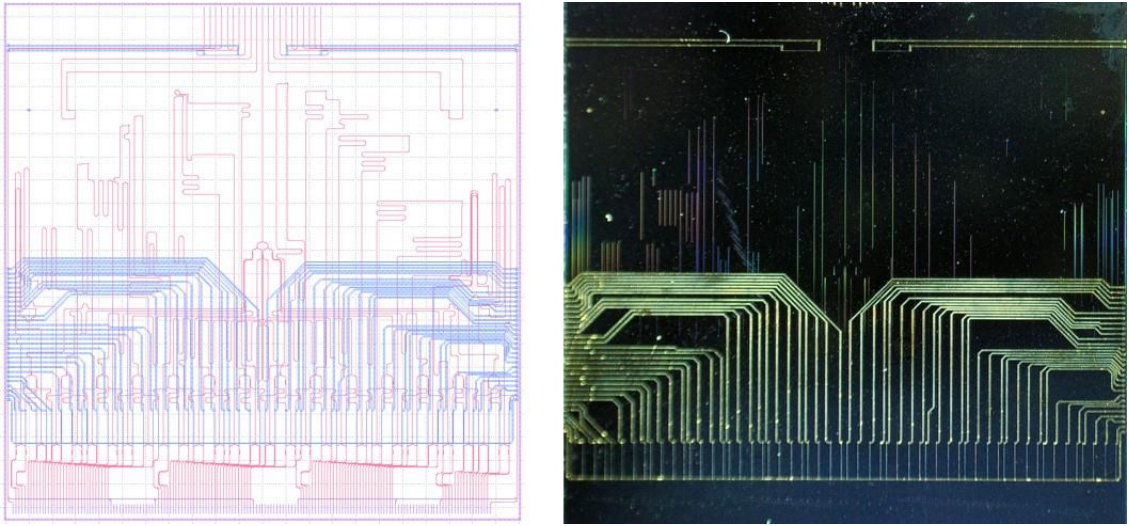

Figure 3. (Left) PIC layout within layout software showing each fabrication layer within the same image. (Right) Microscope photo of the PIC after fabrication and dicing.

**Device Fabrication**

We fabricated the three-layer MICRO PIC using the ASML PAS 5500 300 deep-UV lithography stepper technology. The main fabrication steps are shown in Figure 4 for the $Si_3N_4$ platform including the metal layer depositions. We begin with a brand-new thermal oxide wafer, grown with 6 μm of thermal oxide. For the first layer, we grew 150 nm of stoichiometric $Si_3N_4$ at 765 °C using the Low-Pressure Chemical Vapor Deposition (LPCVD) furnace. The waveguide structure is etched into this $Si_3N_4$ layer using an Inductively Coupled Plasma (ICP) etcher. A 400 nm layer of Low Temperature Oxide (LTO) is deposited at 400 °C in a different LPCVD furnace, then planarized down to the target 50 nm thickness using chemical mechanical polishing (CMP). The deposition process is repeated two more times to stack the three $Si_3N_4$ waveguide layers over each other. Lastly, the top waveguide is covered in 4 μm of LTO which is planarized using CMP. Once the $Si_3N_4$ process is finished, we switch to an electron-beam evaporator to deposit the electrodes. A 700 nm layer of titanium is deposited over the phase shifter regions to act as the heating element for the thermal phase shifting. While other metals such as platinum could be used, the thickness of the titanium would make the phase shifter more resistant to burning out. We cap the heater element with 20 nm of gold to prevent oxidation. Lastly, we deposit 20 nm of titanium and 1 μm of gold for our routing metal which end at bond pads for later integration.

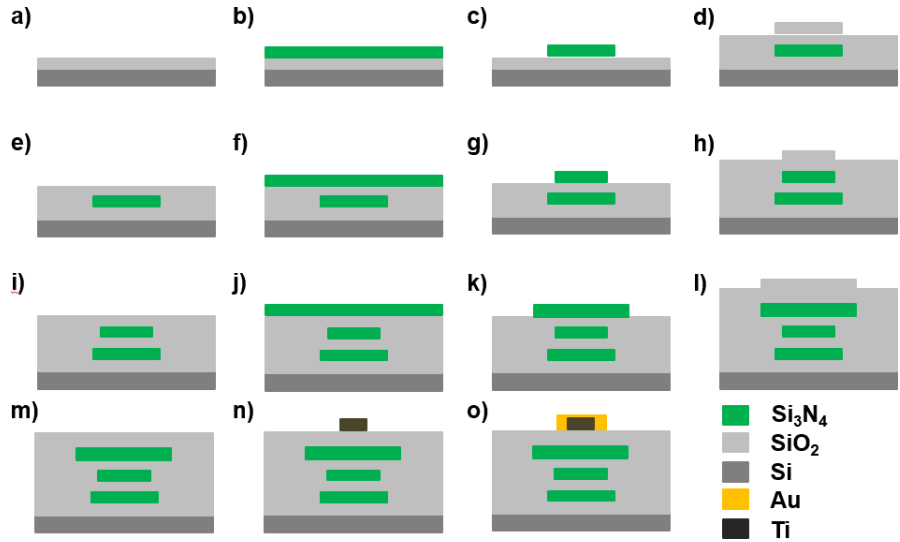

Figure 4. Fabrication steps for the 3-layer Si3N4 PIC including metal traces for heater operation. The Si3N4 layers have a thickness of 150 nm with a vertical spacing of 50 nm filled with SiO2, forming a 3-layer stack for guiding light.

The $Si_3N_4$ waveguides have many design aspects that need to be considered such as device footprint, loss, and functionality. Previously demonstrated in the SPIDER design [T Su, 2018], a 150 nm thick $Si_3N_4$ waveguide layer has been demonstrated to be a good balance between all the factors listed previously. The three $Si_3N_4$ waveguide layers are used to construct the basis for the polarization diversifying gratings used to capture the input signals. To optimize the coupling efficiency of the grating, the spacing between each individual $Si_3N_4$ layers is set to 50 nm. Due to this, the routing of the $Si_3N_4$ is done between the top and bottom $Si_3N_4$ layers to reduce crosstalk that would arise from the waveguide spacing. The waveguide crossing was simulated to have approximately 1 dB of loss per crossing. This can be reduced by adding extra layers of $Si_3N_4$ waveguides along with $SiO_2$ cladding and increasing the separation of the $Si_3N_4$ layers that perform waveguide routing. This allows for the gratings parameters to be adhered to while also reducing potential crosstalk and other inter-layer losses that could arise.

**PIC Characterization**

We measure the propagation loss and bending loss of the $Si_3N_4$ waveguide structures using simple test structures created along the edges of the PIC. For simplicity, we use a tunable laser with the input power measured to be 0 dBm prior to coupling into the PIC. The bending loss is measured by examining a short loopback consisting of two 90° bends and very short straight waveguides, which contribute negligible loss to the loopback structure. The input and output of the loopback are coupled using high numerical aperture (High-NA) fibers to best match the mode of the waveguides at the edge of the PIC. The high-NA fibers are set into a fiber array and edge coupled against the facet of the PIC, incurring a small amount of coupling loss for this setup. The coupling loss using the high-NA fibers are measured to be around 2 dB per coupling interface, which could be partially caused by suboptimal facet polish. Using this value in the loopback structure, the calculated average bending loss from the measured loopback is 0.6 dB/$\pi$ for the $Si_3N_4$ with a bending radius of 100 μm. The optimal radius of the $Si_3N_4$ for low-loss transmission is a radius of 150 μm, however a radius of 100 μm reduces the device size, allowing for tighter PIC routing in place of lower loss.

The propagation loss is calculated by examining the reference structures that run along the perimeter of the PIC. One reference structure has 28.68 mm of straight waveguides while longer structures has 33.02 mm. Each reference structure has 4 bending structures, which are excluded using the results of the bending loss structures to isolate the propagation loss. A final reference structure is observed for a short straight waveguide length of 0.26 cm using one of the looped waveguides in a separate measurement. Similar to the other measurements, the straight waveguide loss is isolated by removing the bending loss. In these measurements, we used lensed fibers to couple light into the PIC due to a change in lab spaces, which resulted in a higher coupling loss of approximately 3 dB per coupling interface. One key difference between the reference waveguides is that the longer waveguide crosses other waveguides due to the limitations on space. The longer waveguide would cross each waveguide perpendicularly to minimize crossing loss. The crossing loss of such waveguide crossings, each with a spacing of 250 nm, are simulated to be approximately 1.1 dB per crossing. By removing this loss, the propagation loss is calculated to be 0.672 dB/cm, which is similar to the losses found in the SPIDER PIC for $Si_3N_4$ of the same geometry [T. Su, 2018]. This can potentially be reduced further by taking steps to smooth the waveguide sidewalls with refining processes such as hot phosphoric acid to remove sidewall roughness.

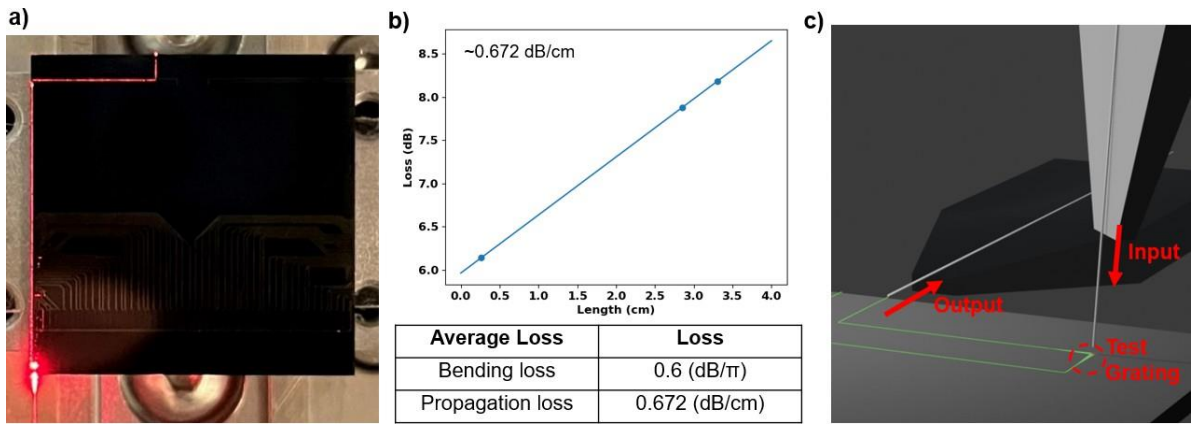

Figure 5. PIC characterization (a) shown with a visible light laser highlighting the reference waveguide. Two reference waveguides are examined to examine is coupling loss and propagation loss (b). Bending loss is measured separately with a separate single structure.(c) Rendered diagram of vertical coupling of fibers for testing the grating reference structure. The Input and output are interchanged to help optimize the input angle into the grating.

On the PIC, there are two grating structures that are used to characterize the coupling loss of the gratings while also helping align other structures, such as lens arrays, to the PIC. To characterize the gratings, a cleaved fiber is vertically coupled into the grating and angled slightly to help couple light in the optimal angle of incidence. An XYZ stage is used to help adjust the height and position of the spot generated by the fiber. We use a cleaved fiber to couple light into the grating and use a lensed fiber to couple the light out of the edge of the PIC. By removing the coupling loss and propagation losses calculated for the other structures, the coupling loss of each polarization on the grating is 6.34 dB and 7.29 dB for S and P, respectively. This translates to about 25% coupling efficiency for the grating, which is similar to the simulated results of the design. The excess loss of the grating results from imperfect matching of the angle of incidence into the grating, resulting in additional loss.

One limitation to this MICRO design is that the inter-layer spacing determined by the grating causes excessive crossing loss. There are 28 different signals coming from the 14 input apertures which are routed from irregular positions all across the PIC. The LO is split into 14 different signals to be shared between each aperture's signals, summing to a total of 42 signals being routed across the PIC. To route these signals, we use the top layer $Si_3N_4$ and bottom layer $Si_3N_4$ to route the signals between input of the gratings to the optical hybrids, using waveguide crossings to allow the signals to travel between waveguide layers. The average propagation distance of each baseline is about 10 cm which results in approximately 6.72 dB of loss. Each baseline goes through an average of $4\pi$ bends, resulting in 2.4 dB of bending loss. The optical hybrid uses multiple 2 x 2 MMI couplers which each have 3 dB loss, plus an additional 0.5 dB loss due to fabrication inaccuracies.

The total theoretical loss from the above analysis is approximately 16 dB from input grating to output. The crossing loss of the waveguide crossings, which are spaced 250 nm vertically, are simulated to be approximately 1.1 dB per crossing. The average number of crossings in this PIC is approximately 10, which increases the optical loss by 10 dB. We continue to work on both the fabrication and design to improve on-chip optical loss by examining potential solutions like increasing the layer count to 5 and utilizing waveguide crossings such as those found in [K. Shang, 2015]. The complexity of the fabrication would increase but the cross-waveguide optical loss could be minimized to ideal values, allowing for measurement of extremely weak signals.

**System Diagram**

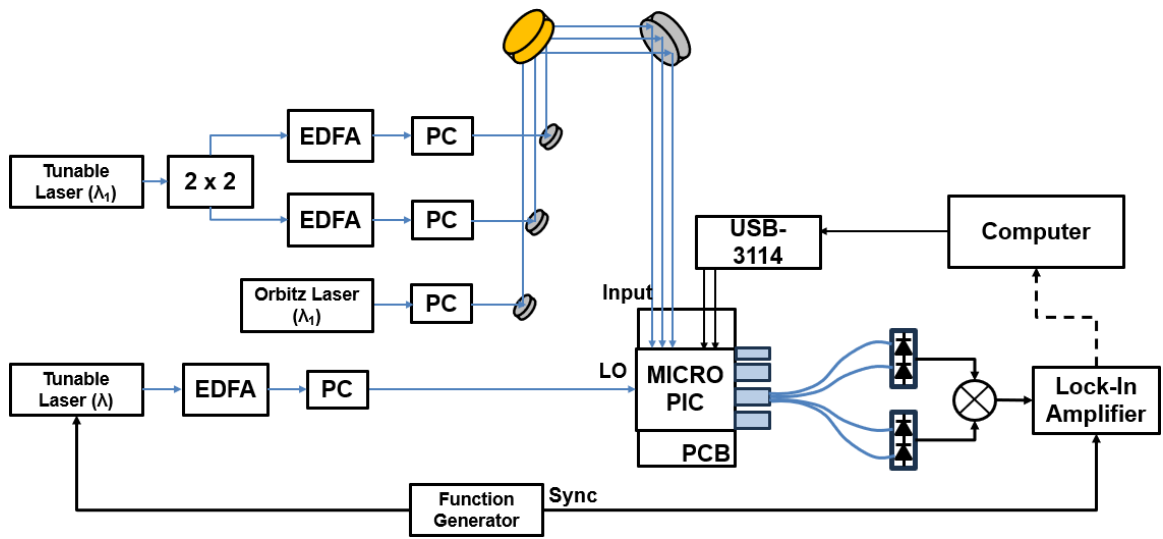

Figure 6. System block diagram for interferometry measurements. Multiple inputs are generated and redirected using mirrors. The mirrors bounce the inputs together into one input image. The input is vertically coupled into the gratings on the PIC while the LO is edge-coupled into the PIC. Fiber arrays are used to acquire the output optical signals and route them to photoreceivers.

In the experimental testbed, show in Figure 6, we use fiber-based optical, sources to generate input signals to propagate them in free-space. We use strong lasers, both tunable and single wavelength, to generate narrow input responses while we use Eribium Doped Fiber Amplifiers (EDFAs) to act as broadband sources. By operating the EDFA without an input, the EDFA generates a broad spectrum based on its operating region, which for C+L band EDFAs it's 1300 nm to 1700 nm. We modify the EDFA spectrum by introducing in-line fiber components, such as tunable wavelength filters, to create unique features in the spectrum. The input signals are collimated using a FL-10B lens and reflected between three mirrors to vertically couple multiple input sources into the MICRO PIC. Each mirror can be adjusted in three angles, allowing the beams to be optimized to the angle of incidence of the gratings on the PIC.

The LO of the PIC is generated by a tunable laser source amplified with an EDFA. The strong laser is edge coupled into the PIC using a lensed fiber where it is split within the PIC using a 1 x 16 power splitter formed by several 1 x 2 power splitters. From the 1 x 16 splitter, 14 output components are routed to each aperture where the LO components are used to perform heterodyne mixing. For each aperture, the LO is split once more and coupled into two 2 x 4 optical hybrids for each input polarization. The optical hybrid combines the input signal with the LO to create two pairs of mixed signals separated by a 90°, producing an I and Q component based on the LO. The optical hybrid outputs are routed off chip to be externally detected using balanced photodetectors.

The output waveguides of the PIC are grouped in batches of 32, where each group of waveguides have a pitch of 127 μm. To match the waveguide pitch, polarization maintaining (PM) fiber arrays with a pitch of 127 μm are aligned to the edge facet of the PIC. These fibers are routed to two PDB410A balanced receivers, where the optical signals are converted to RF signals with a new frequency based on the heterodyned signals. The balanced photoreceivers include transimpedance amplifiers (TIAs) which amplify the photocurrent from the detectors and convert the current to a voltage. The TIA has a gain of 10,000 V/A, allowing for amplification of weak optical signals.

For each baseline, the RF signals are routed to a two-port RF frequency mixer, which multiplies the incoming signals to create two signals that perform frequency addition and subtraction. We use a lock-in amplifier with a low-pass filter to capture and examine the mixed signal at an even lower frequency. This signal is stabilized, integrated, and digitized to be compiled an assembled for imaging on the computer. By combining the frequency mixer and lock-in amplifier, we can use the van-Cittert Ziernike theorem to help extract information of our inputs. One main advantage of using the analog frequency mixing is that the resulting signal can be digitized at a lower speed compared to digital multiplication. This simplifies the complexity of the digitization electronics, allowing for the use of low-power, low-speed analog-to-digital converters (ADCs) which reduces the computational load required to capture the signal. However, the fixed number of ports of the frequency mixer requires the measurements to be taken sequentially during this demonstration. This can be remedied using multi-port multiplication circuits in combination with RF switches to perform the measurements in parallel.

For the spectroscopy measurements, we use the shortest baseline to capture the input light that's projected onto the PIC. By examining the intensity of the shortest baseline, we can reconstruct the input spectrum as we tune the LO and capture unique features depending on the size of the wavelength steps. For each input spectrum, the LO is tuned to start at the shortest wavelength and incrementally stepped through the spectrum, capturing data at each point until it reaches the longest wavelength set for the sweep. Once the sweep is finished, the intensities at each wavelength step is combined together to form the 1-D frequency spectrum.

For 2-D image reconstruction, we measure each baseline of the PIC sequentially to generate 1-D interferograms. Each interferogram is digitized to capture fringe data for the 2-D image reconstruction. For each baseline measurement, one thermal phase shifter in the baseline is electrically biased using a MCC USB-3114 DAC to apply a bias voltage and methodically sweep across range of 0 volts to 6 volts. This bias changes the phase relation between the apertures in the baseline, allowing us to then measure the fringe information needed for image reconstruction. Each baseline's fringe information is combined together into the Fourier transformed source distribution. From that image, we take the inverse Fourier transform to generate the image approximation of the source distribution.

**Results**

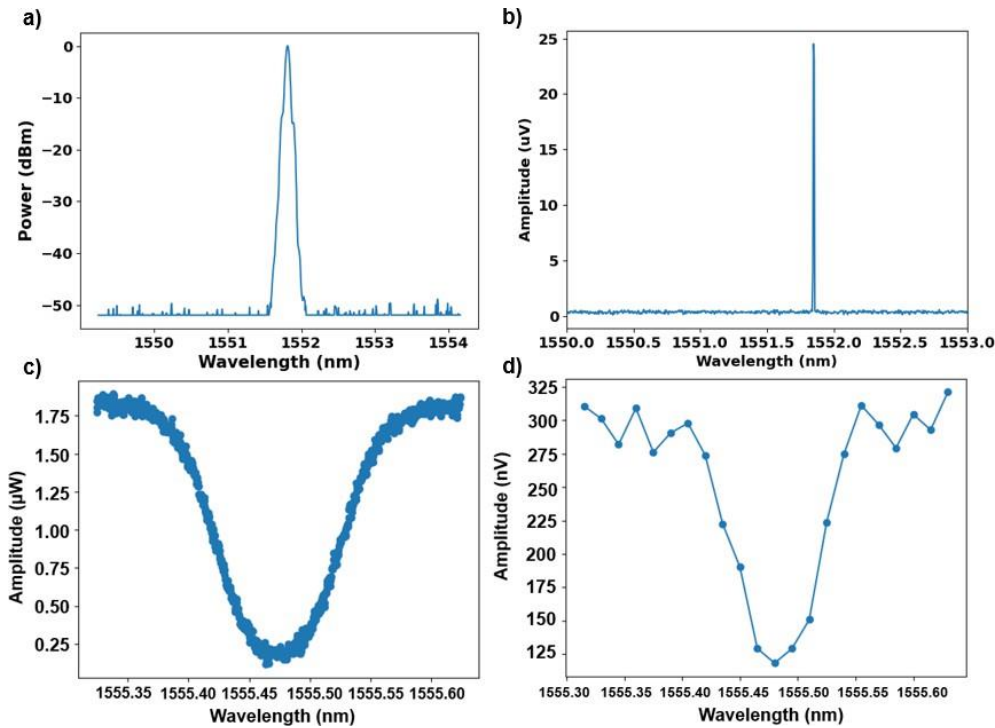

Figure 7. Results of spectrum measurements with different input spectra, including a narrow linewidth laser and a filtered broadband source. a) Shows the optical spectrum analyzer (OSA) spectrum of the narrow-band laser. b) The results of the spectrum recovery after tuning the LO to sample across the wavelength spectrum. c) OSA spectrum of a broadband spectrum that's filtered with a fiber-Bragg grating (FBG). d) Recovered spectrum of FBG filtered spectrum.

For the 1-D spectroscopy measurement, we use the shortest baseline to capture different unique spectrum based on the system in Figure 6. In Figure 7a, b, we couple a strong, single wavelength laser into the input and collimate the beam onto the PIC. The laser has a linewidth of 30 kHz and an output power of 13 dBm, making the spectrum unique. We capture the input spectrum on an optical spectrum analyzer (OSA), shown in Figure 7a, as a reference to compare our recovery results. For spectrum recovery, the LO is initially tuned to 1550 nm and tuned across a 5 nm wavelength range in steps of 0.001 nm. When comparing the results, the OSA plot has a broader spectrum compared to the recovered results in Figure 7b. This is due to the resolution limit of the OSA, which is locked at a minimum of 0.02 nm. The resulting reconstruction creates a delta function at the input laser's wavelength.

Then, we replace the input with a broadband source that is modified with a unique spectrum. For the source, we use a C+L band EDFA with no input to act as an arbitrary spontaneous emission (ASE) source with a wavelength range of 1300 nm to 1700 nm. A Fiber-Bragg grating (FBG) is placed at the output of the EDFA to filter the spectrum with a notch response. The FBG has a center frequency of 1555.473 nm with a full-width half-max (FWHM) of 0.1 nm. The filtered spectrum is captured in the OSA, shown in Figure 7c, where the dip of the FBG is captured. Since the FWHM of the FBBG is 0.1 nm, the LO laser is tuned across a range of 0.3 nm to span the entire range of the FBG. Each wavelength step is 0.025 nm across the tunable range to ensure a sufficient number of points were in the FBG dip. The recovered spectrum, shown in Figure 7d shows the FBG spectrum notch approximately centered at the original specified center wavelength. The extinction ratio of the recovered dip is approximately 5 dB, measured from the dip to the edges of the FBG area. The OSA's plot shows the extinction ration is approximately 8.5 dB. This likely due to the measurements of the reconstruction reaching the noise floor of the system, which would limit the depth of the FBG.

Finally, we expand the measurements to include various baselines across the PIC. The input is replaced with a laser to generate a simple point source function (PSF) which is directed onto the PIC. The laser mode profile is expanded using a FL-10B lens and collimated onto the PIC at approximately 2° from normal. In this case, the PSF is simulated as a narrow gaussian beam, where the Fourier transform of the beam is a gaussian of a slightly different size shown in Figure 8a. This beam acts as a pseudo-stellar body in our experimental setup to simulate the universe. Each baseline in the PIC gathers fringe data unique to the baseline positions. The reconstructed image is shown to have peak intensities based on the simulated locations of the inputs. The reconstructed image appears to be sparsely sampled, due to the limited uv-plane coverage of the existing baselines on the PIC. This can be solved by rotating the system to change the positions of the aperture in the image plane, gathering more unique data points. In addition, we can use multiple PICs from a wafer to scale the number of baselines formed and acquire a large number of samples.

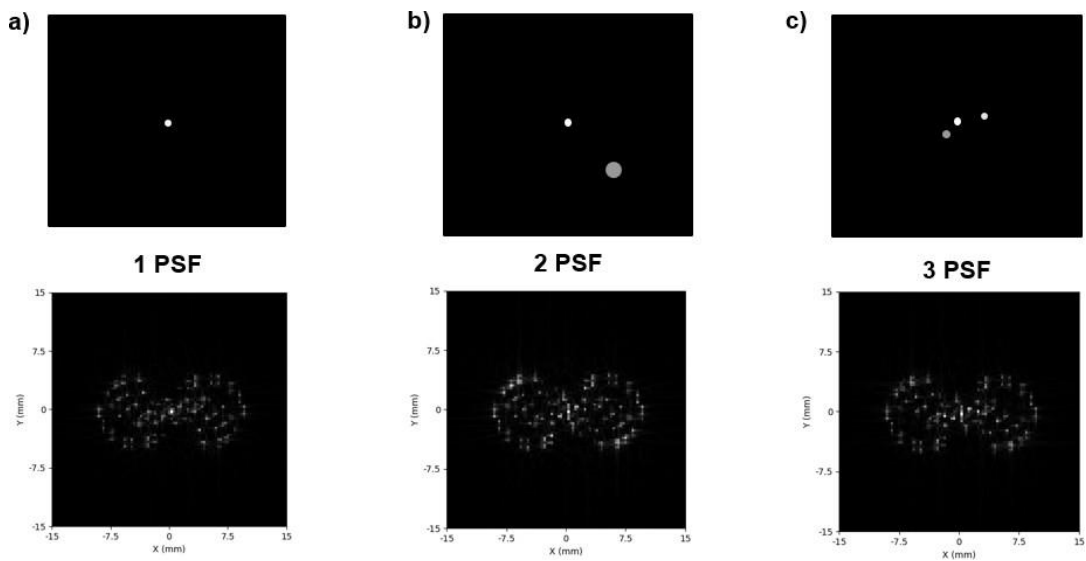

Figure 8. 2-D Image reconstruction with increasing number of inputs. In the 2 PSF case, one input is centered at the center of the PIC while another input is spaced 14 mm southeast of the center. In 3 PSF case, one input is placed in the center while another input is spaced 3 mm east and another is spaced 1 mm southwest.

In Figure 8b, we demonstrate simple image reconstructions of the three different sets of input sources. We position multiple inputs at varying distances across the PIC to create unique inputs. For the two input measurement, a collimated laser source with a 1 mm beam diameter and 13 dBm power is placed in the center of the array. Another laser source with 6 dBm power is collimated to have a beam size of 2 mm positioned 14 mm south east of the first source. In Figure 8, we find that the center is illuminated with the strong source, while a more diffuse response is found southeast of the center point. This response is mirrored weakly on the other side of the image due to the sample mirroring of the complex visibility.

In Figure 8, the center is illuminated with the 13 dBm laser source which is collimated to a beam size of 1 mm on the PIC surface, centered within the PIC area. A laser source is amplified to 10 dBm and is collimated to 1.5 mm positioned 3 mm east to the first source. A third source is amplified to 6 dBm and is collimated to 0.5 mm positioned 1 mm southwest of the first source. This creates three unique PSFs positioned closely to each other around the first source, mimicking a cluster of stars. We find that the 3 points are visible around the center of the reconstructed image.

Each reconstruction suffers from two main problems: PIC loss and sparse sampling. In our design, the restriction of layer gaps imposed by the grating design greatly increases the loss in routing the signals across the PIC. This reduces the signal quality captured by each baseline, inhibiting the image quality. We can mitigate this problem by modifying the PIC design to reduce the losses across each optical component. In addition, each baseline generates a small signal even if they capture little of the input signal. This means that the background noise, whether from other light sources or the detector noise, contributes to the image, generating noise and reducing image quality.

**Conclusion**

In this paper, we discuss the MICRO imaging system design for creating a heterodyne interferometry system to reduce SWaP. We demonstrate a 14 aperture $Si_3N_4$-based PIC for applications in spectroscopy and interferometric imaging. The PIC can observe unique spectral features with a single baselines with a far-field beam. The imaging testbed shows the 2-D image reconstruction of a simple input source using the PIC for measurement and reconstruction. This experiment demonstrates the basic principles of the MICRO imaging design. Future iterations of the PIC would modify the fabrication process to reduce the optical loss on the PIC. The changes in the fabrication process would help to improve system sensitivity to weak input signals. In addition, the system would include rotation plates to shift the position of the baselines, providing better coverage of the uv-plane and filling in holes of coverage. We plan to examine methods of integrating on-chip photodetectors and interface technologies, such as electronic to photonic interposers, to acquire the photocurrents directly from the chip. Furthermore, the implementation of simultaneous measurement schemes using analog method, digital methods, or a combination of both will be explored. All of these would work to create an enclosed system capable of performing real-time image reconstruction in a compact system. Future work to image the sun would add a quarter waveplate to the front of the PIC, converting selected circularly polarized light into linearly polarized light for capture on the PIC. In addition, standard algorithms such as CLEAN can be used to further improve the image reconstruction.

**Funding Sources**

National Aeronautics and Space Administration (NASA) Grant # 80NSSC20K0321

**ACKNOWLEDGMENT**

The authors would like to thank technical support and assistance from the Marvell Nanolab at University of California, Berkeley and Center for Nano-MicroManufacturing at University of California, Davis.

**Author Contributions**

The manuscript was written through contributions of all authors. All authors have given approval to the final version of the manuscript. ‡These authors contributed equally. (match statement to author names with a symbol)

ABBREVIATIONS

PIC – Photonic Integrated Circuit

LO – Local Oscillator

SiPh – Silicon Photonics

SWaP – Size, Weight, and Power

$Si_3N_4$ – Silicon Nitride

$SiO_2$ – Silicon Dioxide

LPCVD – Low Pressure Chemical Vapor Deposition

CMP – Chemical Mechanical Polishing

FWHM – Full Width Half Max

PSF – Point Source Function

EDFA – Eribium Doped Fiber Amplifier

ICP RIE – Inductively Coupled Plasma Reactive Ion Etch